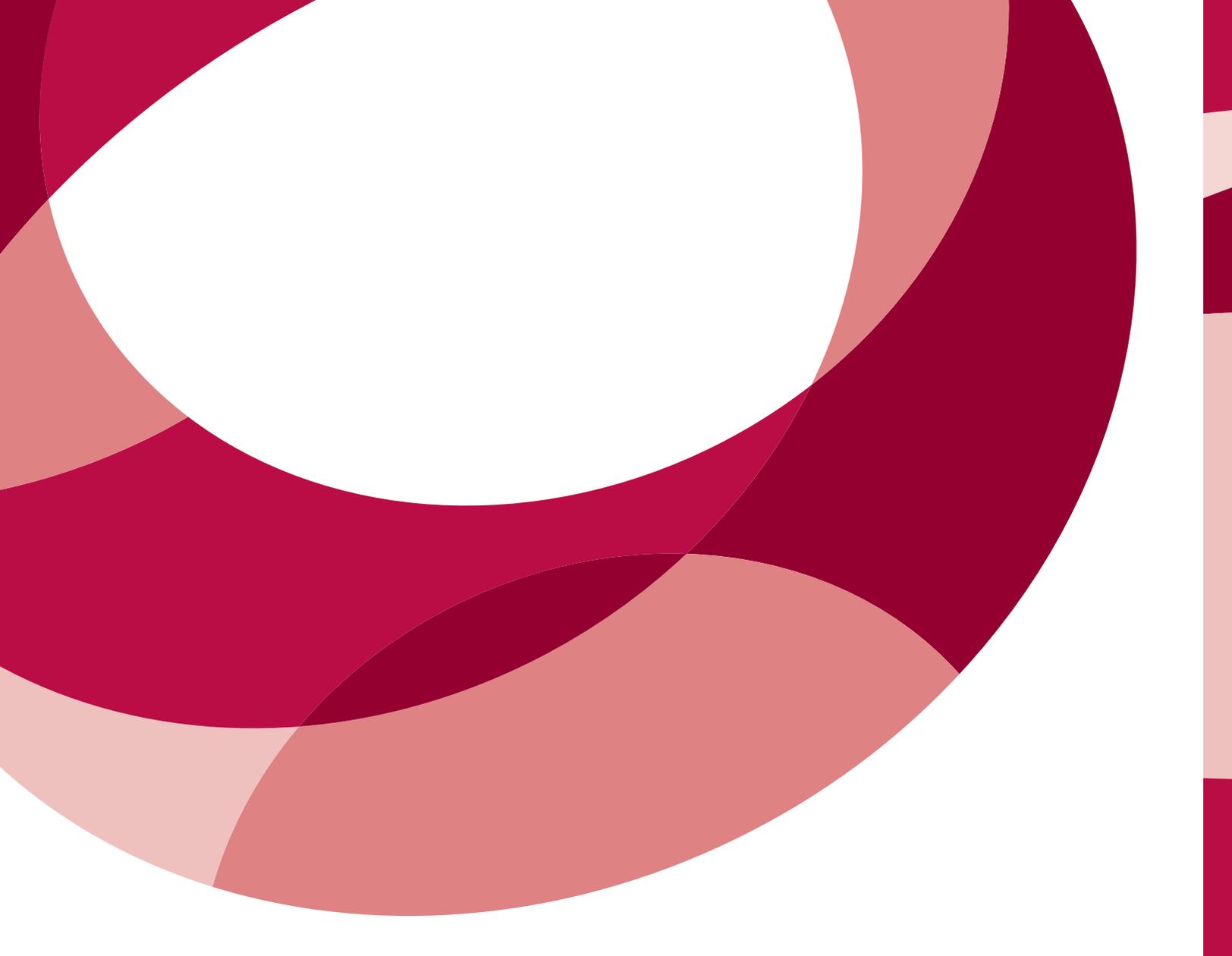

# A New Age of Computing and the Brain

REPORT OF THE CCC BRAIN WORKSHOP

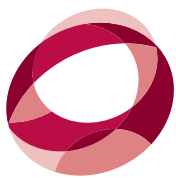

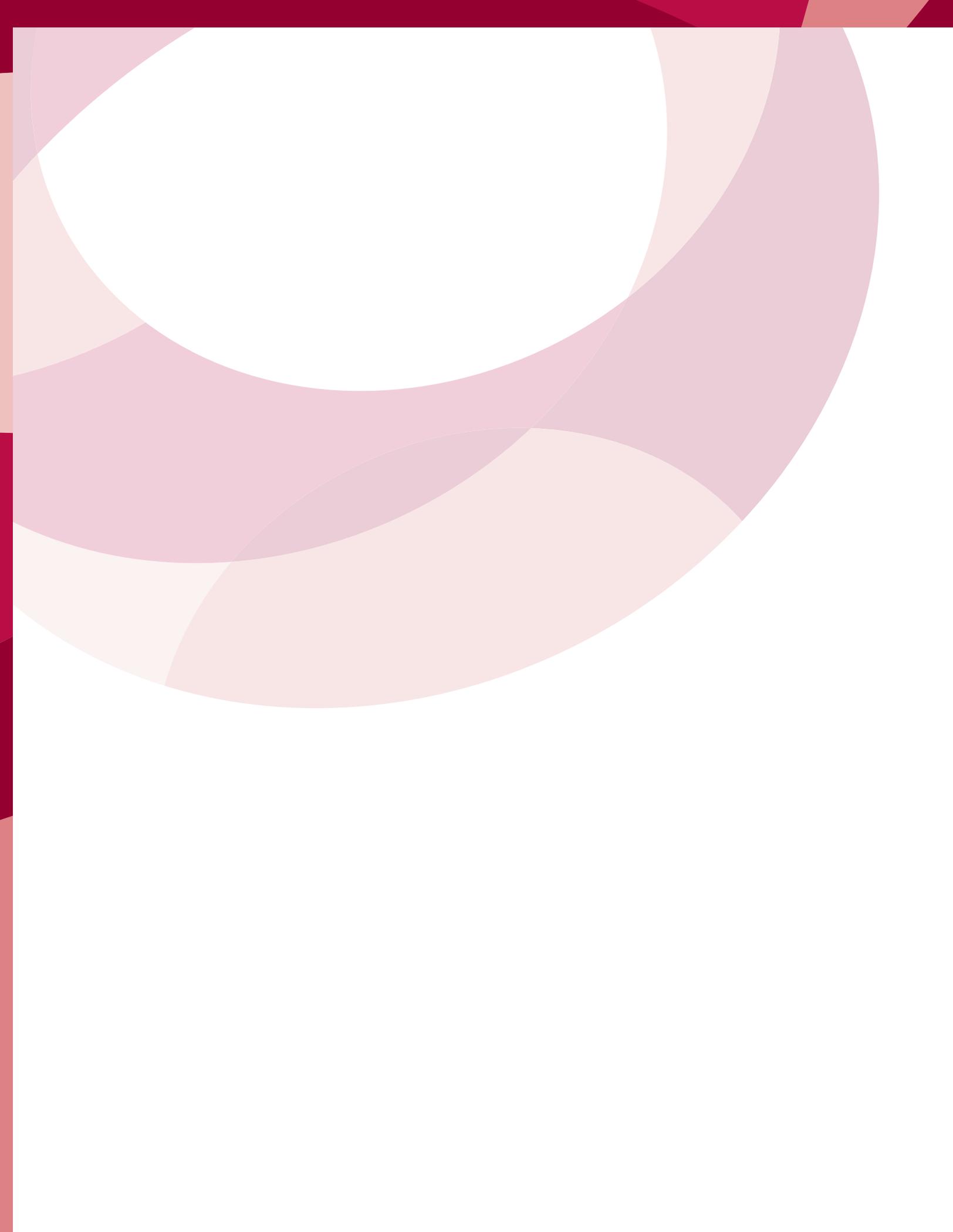

# A New Age of Computing and the Brain

Including contributions by:

Polina Golland, Jack Gallant, Greg Hager, Hanspeter Pfister,

Christos Papadimitriou, Stefan Schaal, Joshua T. Vogelstein

Sponsored by

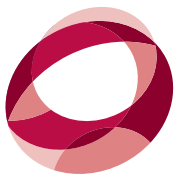

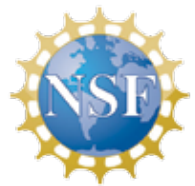

This material is based upon work supported by the National Science Foundation under Grant No. (1136993). Any opinions, findings, and conclusions or recommendations expressed in this material are those of the author(s) and do not necessarily reflect the views of the National Science Foundation.

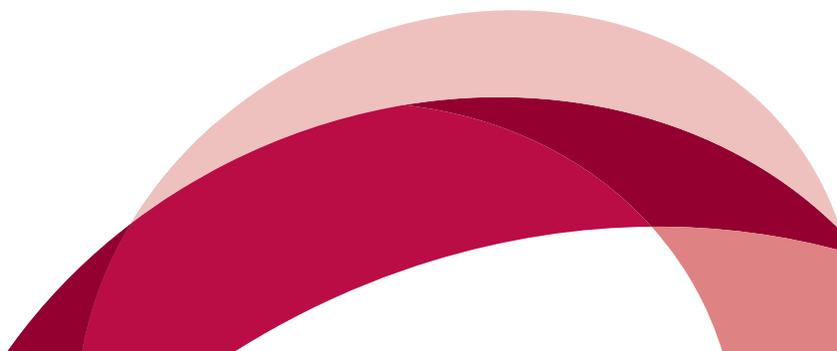

## I. Overview and Summary

Throughout its history, humankind has been fascinated by a question that is simple to pose, yet remarkably resistant to resolution: "How does the brain work?" Philosophers have debated the workings of the mind for centuries. Da Vinci made detailed sketches of the brain. By the turn of the century, scientists began to understand some of the brain's basic structure and function. Today, we can image and record brain activity from the neural to whole-brain level. Yet, divining how the structure and function of the several billion neurons and their trillions of interconnections leads to the complexity, diversity, and adaptability of human behavior continues to elude us. It is indeed ironic that almost every advance in brain science has given us a deeper appreciation of the challenges of understanding the brain.

The history of computer science and brain sciences are intertwined.[1] In his unfinished manuscript "The Computer and the Brain," von Neumann debates whether or not the brain can be thought of as a computing machine and identifies some of the similarities and differences between natural and artificial computation.[2] Turing, in his 1950 article in Mind, argues that computing devices could ultimately emulate intelligence, leading to his proposed Turing test.[3] Herbert Simon predicted in 1957 that most psychological theories would take the form of a computer program.[4] In 1976, David Marr proposed that the function of the visual system could be abstracted and studied at computational and algorithmic levels that did not depend on the underlying physical substrate.[5]

Today, we stand at a point where exponential advances in the science and technology of computing and concomitant advances in approaches to brain sciences have ignited new opportunities to forge connections between these two fields. Many of these opportunities were not even on the horizon as little as 10 years ago. Consider the following:

◗ Data related to brain research has exploded in diversity and scale, providing unprecedented resolution of both anatomy and function across a growing population of individuals, but new challenges for brain sciences. EM Connectomics, MR Connectomics, and functional imaging are but a few of the growing number of examples.

◗ Access to enormous computational power coupled with computational data science tools has been revolutionized by the growth of cloud-based computing platforms. Other sciences such as astronomy and genomics have already successfully exploited these new resources. Brain science can be the next "big data science" to create a new computational lens through which to study, and connect, the structure and function of the brain.

◗ The surprising success of new models for machine learning inspired by neural architectures is reigniting directions of inquiry on biomimetic algorithms. These successes also will begin to provide insights and inquiries that may influence our thinking about the brain, how it may function, and how to test those ideas.

◗ New methods for acquiring and processing behavior data "at scale" are emerging from the mobile device revolution, providing new possibilities for brain scientists to connect behavior to function to structure in ways that were heretofore impossible.

---

[1] For the purposes of this document, we will use the term "brain sciences" to represent all disciplines that contribute to our understanding of the human brain, including neuroscience, cognitive science, brain imaging, psychology, and other neural and behavioral sciences.
[2] J. von Neumann, *The Computer and the Brain*, Yale University Press, 1958.
[3] A.M. Turing, "Computing Machinery and Intelligence." [online]. Available: http://www.csee.umbc.edu/courses/471/papers/turing.pdf. [Accessed: April 20, 2015].
[4] A. Newell and H. A. Simon, "Computer Science as Empirical Inquiry: Symbols and Search." [online]. http://dl.acm.org/citation.cfm?id=360022. [Accessed: April 20, 2015].
[5] T. Poggio, "Marr's Approach to Vision." [online]. Available: ftp://publications.ai.mit.edu/ai-publications/pdf/AIM-645.pdf. [Accessed: April 20, 2015].

This presages a day when traditional laboratory science meets "data in the wild" acquired in less controlled real-world situations, creating new opportunities and challenges for data analysis, hypothesis testing, and behavioral modeling.

These are just a few of the opportunities that lie ahead if we can develop a dialog that creates synergistic partnerships between computer science and brain science.

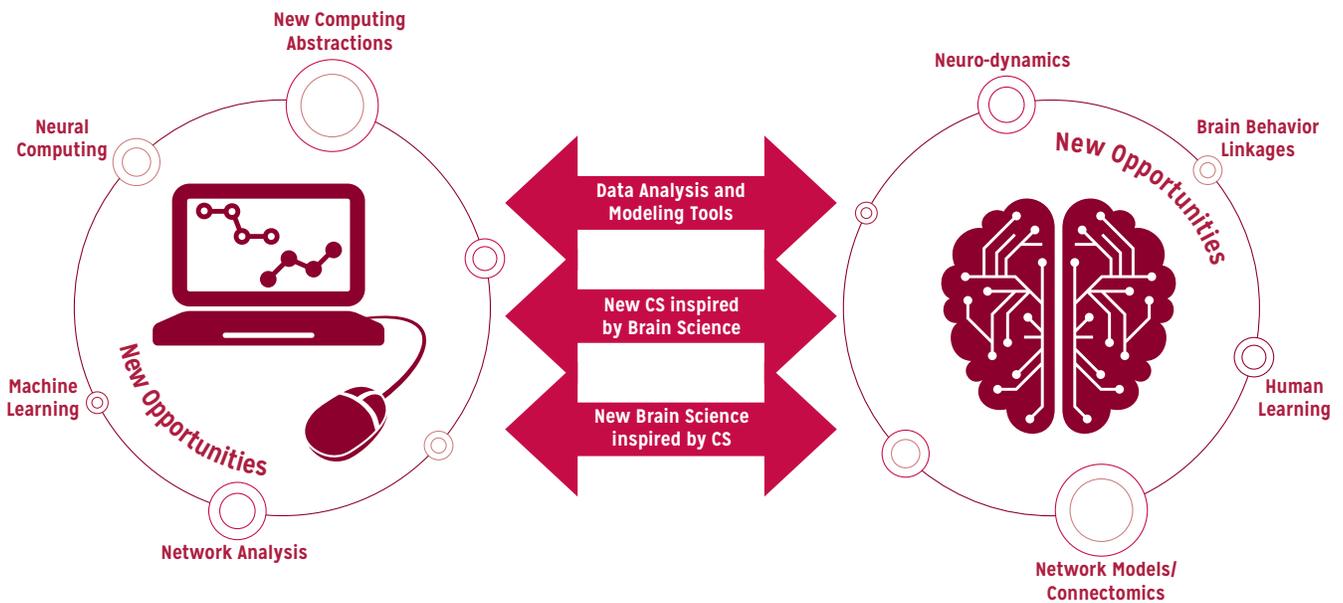

*Figure 1: Advances in computing and brain science will require increased access to data and analysis tools, the development of new computing concepts to advance brain science, and will ultimately lead to new insights that will advance computing as well.*

In December 2014, a two-day workshop supported by the Computing Community Consortium (CCC) and the National Science Foundation's Computer and Information Science and Engineering Directorate (NSF CISE) was convened in Washington, DC, with the goal of bringing together computer scientists and brain researchers to explore these new opportunities and connections, and develop a new, modern dialogue between the two research communities.

Specifically, our objectives were:

1.  To articulate a conceptual framework for research at the interface of brain sciences and computing and to identify key problems in this interface, presented in a way that will attract both CISE and brain researchers into this space.

2.  To inform and excite researchers within the CISE research community about brain research opportunities and to identify and explain strategic roles they can play in advancing this initiative.

3.  To develop new connections, conversations and collaborations between brain sciences and CISE researchers that will lead to highly relevant and competitive proposals, high-impact research, and influential publications.

The workshop was designed to drive an effective dialogue around these objectives. Speakers and panelists focused on carving out well-delineated and high-impact problems that could be further discussed and articulated in the breakout sessions, focusing on these themes:

◗ What are the barriers to progress in cognitive, behavioral, or neuroscience that would be targets of opportunity for CS-research? Where is genuinely new computer science needed, and where are new computational ideas created?

◗ What are areas of computing research that might benefit from or be informed by brain sciences?

◗ How can the connections between the two fields be enhanced through the development of common computational and data tools, and analysis methods?

◗ Are there grand challenges in this interface – big ideas that are well scoped, easily communicated, and where progress will be quantifiable?

Four broad topical areas were explored through panel discussions: (1) brain mapping, (2) connecting brain, mind and body, (3) challenges in data, and (4) opportunities in computing. Each pointed toward new challenges and opportunities where computer science and brain sciences could collaborate.

1. **Brain Mapping**: New imaging modalities are dramatically increasing the resolution, scale, and volume of brain imaging data. At one end of the scale, we can observe neuroanatomy at nanometer resolution; at the other we can now observe whole brain functional behavior over extended periods and under a variety of stimuli. Extracting meaningful information from this diversity, complexity, and scale of data can only be done via advanced computational tools. What are those tools? How can they be scaled as images continue to grow in size and complexity? How can correlative information be extracted from images of different kinds? Can we build predictive models that can relate stimuli to brain function? And at a higher level, what sort of infrastructure, training, and access mechanisms are needed to enable this type of research?

2. **Connecting Brain, Mind and Body:** Studies of behavior are also becoming increasingly quantitative and data driven. In particular, studies of motor behavior, the visual system, hearing, speech, touch, taste and smell continue to advance and inform us of individual "subsystems." Cognitive science and psychology study higher-level questions, such as "how do we learn," "what is our memory capacity," or "what do we attend to". But, these are incredibly diverse and complex questions. How do we "connect the dots" through theories and methodologies relating low-level mechanisms of neural computation to more abstract information processing and systems to high-level behavior? What computational tools and environments are needed to support replicable, scalable science extending beyond a single experiment or a single laboratory? How do we incentivize scientists from all parts of brain sciences to contribute to and to unite behind these efforts?

3. **Challenges in Data:** At the heart of the brain/computing research interface lies the great challenges of the volume, velocity and variety of brain sciences data. The challenge is to relate the many different scales and modalities of data in ways that will support new kinds of scientific collaboration. Data must be linked across scales, modalities, and experiments. Perhaps most importantly, data must be democratized. Just as the World-Wide Web created an unprecedented environment of data wealth available to all, so must brain sciences data be made universally available – by aggregating the "long tail" of data collected by every investigator in the country, and by creating tools to share, analyze, visualize and compare all forms of data connected to brain sciences.

4. **Opportunities in Computing:** It is unquestionable that advances in brain sciences will require new ideas in computer science. Much like genomic analysis, the raw complexity of the problem can only be tamed by creating computational tools and computational models that support both the statement and testing of fundamental scientific hypotheses. Detecting patterns in neural architecture and neural behavior at scale is a massive problem in data mining and modeling. Computer simulations will be reference models for testable hypotheses. Conversely, understanding fundamental neural computational principles may open the door to new understanding of algorithms for learning, adaptation, as well as catalyzing new directions toward creating devices that assist humans in intuitive, synergistic ways.

The report concludes with a future vision for brain science that is both driven by, and informing to, advanced computing research.

## 2. Brain Modeling

Scientists are recording the activity and mapping the connectivity of neuronal structures to help us understand how brains work. Computational analysis and modeling promises to extract relevant information from the tremendous amounts of measurement data. While recent advances in measurement technology have enabled us to map various aspects of the brain architecture and function, many great challenges remain. Going forward, technological innovations at the interface of computing and neuroscience will benefit greatly from computational methods to extract information from available and future data consisting of images of the brain structure and signals of brain activity.

Currently, several measurement modalities exist that provide a glimpse into the architecture of the brain at vastly different scales. Each modality offers its own advantages, but we have yet to identify a robust approach to answering detailed questions suggested by theories of brain function, or to proposing new hypotheses on brain organization from available measurements. Electron microscopy (EM) provides an exquisite level of detail, all the way down to synapses, but for very small volumes; diffusion and functional MRI offer low-resolution (in space and in time) images of the entire brain; M/EEG yields high temporal resolution of the brain signals but with poor spatial localization.

As new brain mapping technologies become available, computational analysis and modeling promises to bridge the gap between theories of neuronal systems in the brain and measurements of these systems that are feasible to acquire. Moreover, many brain mapping experiments produce vast amounts of data that must be analyzed to extract concise models of brain organization and function. This necessitates development of computational tools that can handle large amounts of data and support modeling and algorithmic developments in the field. Finally, the field is clearly in need of computational infrastructure that facilitates and encourages sharing of data and analysis methods.

**Data Analysis and Modeling:** One of the biggest barriers to mapping the architecture of the brain are methods to *identify general and local (system-specific) motifs and patterns* in the acquired data (both images and signals). This is important at all scales of imaging and requires new computational methods in feature detection, classification, data management, visualization, and analytics. We need new analytic approaches to discover common structure and organizational principles in patterns of functional and anatomical connectivity. And it will be important to model imaging data jointly with other types of information available to us. This includes modeling at different scales, from genetic influences of cell function to discovery of genetic influences on global connectivity patterns.

To really understand the brain structure, function, and their interconnections will require multimodal data fusion with the goal of extracting a model based on more than one imaging modality. Examples include integration of EM and light microscopy for synapse detection and modeling, integration of M/EEG with fMRI to create descriptors of activation patterns at better spatial and temporal resolution, and relating optical Calcium imaging to fMRI blood-oxygen level data. To integrate this data requires common reference frameworks (atlases) to enable multimodal data fusion across scales, including frameworks that identify functional alignment, not just anatomical alignment across individuals.

A NEW AGE OF COMPUTING AND THE BRAINIn addition to new methods for handling large image and signal datasets, we need new theoretical methods and computational structures to handle very large-scale data sets. This includes new graph theoretical models for analysis, methods to discover and analyze large-scale connectivity and function in dynamic graph structures, and methods for describing and finding motifs in heterogeneous geometry and graph structures.

The brain is a complex, nonlinear dynamical system, and there are likely deep connections between the architecture of the brain and its dynamic behavior. Thus, understanding brain motifs not just from an architectural point of view, but also from the perspective of dynamical behavior will be essential to gain insight into brain function. Current methods for measuring brain activity tend to suppress dynamical signals (EEG is low-dimensional and fMRI is slow), but future developments in instrumentation will dramatically increase the amount of dynamical information that can be recovered from brain measurements. At present there are few methods for analyzing and modeling complex dynamical systems with limited data. Improved frameworks for dynamical systems analysis would improve our ability to understand, model and predict brain function enormously.

Finally, we will need new statistical approaches that improve inference and prediction accuracy in complex neuroscientific datasets. In particular, we must develop better methods for modeling individual variability and for accounting for measurement and statistical error in the data. For example, statistical methods for testing hypotheses on neural graphs, or methods for assessing the similarity of dynamical models recovered from recordings of brain activity will be needed. Better methods for identifying and reducing the impact of false positives and false correlations, which are inevitable in large data sets, are also needed.

**Computational Tools and Infrastructure:** The need to computationally identify motifs and patterns and the accumulation of ever-greater amounts of brain data poses major infrastructure challenges for computer science. On one hand we need better data repositories and structures to facilitate management (storage, curation, sharing) of heterogeneous large-scale data acquired in diverse brain mapping experiments. At the same time, we need tools for large scale processing that are well suited to brain data. In contrast to many other data-intensive fields that produce vast numbers of small data records (finance, geomodeling, etc.), brain mapping problems involve large collections of large observational elements. This motivates the development of different type of computational infrastructure and different (often data specific) data access methods than what has been developed for many other "big data" applications.

In addition to new methods for handling large image and signal datasets, we need new theoretical methods and computational structures to handle graph and geometric data. This includes new graph theoretical models for analysis, methods to discover and analyze large-scale connectivity and function in dynamic graph structures, and methods for describing and finding motifs in heterogeneous geometry and graph structures.

Many of the graph theoretic concepts and techniques used in neuroscience currently were actually developed in the 1990s in response to the advent of the Internet. These approaches aren't really optimized for representing biological data, which have very different signal and noise properties. Furthermore, once large-scale connectivity and functional information are available, new computational systems (software and hardware) to simulate large-scale networks based on theoretical models of neuron populations will be required. All of this requires optimization methods and supporting computational infrastructure to perform large-scale parallel computation for extracting computational models from data.

## 3. Connecting Brain, Mind, and Body

The study of "Brain, Mind, and Body" recognizes the importance of grounding studies of the brain in the physics of bodies and sensory information, in the behaviors that are to be generated by a living being, as well as in the information processes that can facilitate cognitive processes. These studies make connections between brain sciences and topics of cognitive psychology, computational motor control, artificial intelligence (AI), machine learning (ML), and robotics. In some sense, a classical claim by Richard Feynman is addressed in that "What I cannot create I do not understand".[6] AI, ML, and robotics are largely about synthesizing systems that accomplish some form of autonomy and competence in real world tasks. The information processing in these tasks is often based on information emitted from a physical environment, similar to what has to be processed by living beings. Action with a physical body becomes the evidence of successful behavior, and many components of physics in biological and synthetic systems are the same. Cognitive psychology, computational motor control, and related fields address the computational interface between synthetic and biological systems, maybe something that could be called the level of "theory and algorithms" in the spirit of David Marr. But it is often difficult to connect this more abstract computational thinking to neuroscientific data that are measured at the implementation level. What is the role of these more abstract theories of information processing in brain sciences? How can collaborative projects be initiated, how can they be successful? How can it be assured the US BRAIN initiative is not only about low-level observations of the brain, but rather also connects to behavior and higher level information processing?

**Computational Theories and Models:** The presentations during the workshop illustrated such problems from four different viewpoints, all inspired by behavior and/or computational theories, and all interested in how more top-down thinking can connect to the low level mechanisms of information processing in the brain. For instance, when observing motor behavior, how can one discover structure in this behavior, the decision-making processes behind it, and the intent of the behavior (often formalized as an optimization criterion). Computational theories address such processes often in terms of reinforcement learning. The current wave of "deep learning" has some interesting mechanisms to discover structure without making many explicit assumptions, but the domain of behavioral analysis is not really addressed by deep learning research yet.

Another topic is how computational approaches can be used to develop models to understand cognitive processes of the mind, for instance, as in language processing and speech production. While some ideas for such high level cognitive computational models exist, their connection to neural implementation remains rather vague. Another example of computational thinking revolves around the question of how the brain processes uncertainty. The representation of uncertainty has been a long-standing question in computational neuroscience, and there are currently insufficient data to constrain existing models.

Finally, there is an overarching question as to whether or how biological data can lead to normative theories that can be verified in experiments. That is, what constitutes a computational model of a biological process, and what level of measurement and testing constitutes verification? What level of mechanism should a model expose? A "black box" that is able to replicate some type of "input-output" behavior (e.g. predict a measureable subject response from a visual signal) is intriguing and highly measurable, but does it provide real insight? A model that is built on a complex neural model may seem more valuable, but how would (or should) one measure, compare or verify that the artificial neural structure is consistent with brain activity?

**Future Research Directions:** the current brain initiative focuses largely on collecting low level and detailed data of the brain, with little emphasis on behavior and computational theories. The discussions above indicate some ambiguity as to how behavioral and cognitive research can be connected directly to this initiative. As a middle ground, suggestions were developed that a "Big Data" approach for motor science, cognitive science, and behavioral science could provide a foundation, which could enable future research developments towards a combination of low level brain sciences and behavioral sciences. The methods needed to collect, process, and distribute such data provide several significant challenges.

---

[6] "What did Richard Feynman mean when he said, "What I cannot create, I do not understand"?." [Online]. Available: http://www.quora.com/What-did-Richard-Feynman-mean-when-he-said-What-I-cannot-create-I-do-not-understand. [Accessed: April 20, 2015].

*What Kind of Data?* One critical issue is what data one should actually try to collect for neurobehavioral studies, and what efforts are associated with this data collection. Three complementary options present themselves:

1. **Experimental Data:** Design experiments to measure as complete data as possible, for instance, including movement data, eye movements, interaction forces with the environment, physiological measures, EMG, EEG, environmental data and context, social context, etc. The required instrumentation would be quite significant and costly, and would most likely resemble an "intelligent house" for neuro data collection.

2. **Found Data:** Collect as much data as possible, from cell phones, wearable devices and the Internet. While highly incomplete, such data, when collected massively, may allow uncovering hidden state and discovering interesting issues. As an analogy, fMRI data provides a very coarse and incomplete snapshot of brain processing, but nevertheless, has allowed us to make numerous discoveries.

3. **Simulated Data:** Create elaborate simulators that can create essentially endless amounts of data, and which would serve as a test bed for what data is actually useful to collect.

*Data Analysis:* Assuming the existence of such data repositories, depending on the realm of interest, many different goals for data mining could be pursued, including general data mining and structure detection in neurobehavioral data. Among those, one can distinguish between functional analyses and clinical analyses.

1. Functional analyses aim to uncover principles of information processing of the brain. Those include theories of optimization (optimal control, reinforcement learning), the extraction of a formal description of behavioral intent, biometrics and variability of behaviors, decision making processes, behavior prediction, emotional analyses, etc.

2. Clinical analyses aim to detect diseases, potential risks (e.g., as in early detection of oncoming dementia), correlations of behavior with quality of life, correlations of biometrics with clinical conditions, etc.

Big neurobehavioral databases could make a significant contribution to the US brain initiative, and will prove critical in the effort to establish a data-driven understanding of the brain. Creating such data repositories presents formidable challenges in the forms of instrumenting people and the environment, data structuring, and data interpretation. The goal should not merely be to produce a descriptive analysis of the data, but rather to support a functional analysis of data collected under ecologically valid conditions. Simultaneous collection of detailed behavioral data with advanced neurophysiological measurement would provide an unprecedented opportunity to integrate low-level neuroscience with behavioral and cognitive neuroscience.

## 4. The Challenges of Data

As already noted in the previous sections, data generation in brain sciences has rapidly accelerated, leveraging research advances in genetic, molecular, cellular, imaging, and electrophysiological approaches, among others. However, tools for systematically archiving, integrating, and disseminating data generated through divergent experimental techniques lag far behind. The CISE community can certainly contribute to developing open-science platforms that enable large, heterogeneous data sets to be combined and exploited to develop increasingly detailed and comprehensive models of neural function (Figure 2).[7]

The overarching observation is that neuroscience is the next science to get to big data, following astronomy and genetics. We define big data in this context as any data that breaks the contemporary scientific workflow dogma. In neuroscience today, that dogma is that a single experimentalist collects data, that data is stored into local storage (e.g. a hard drive), and then analyzed via loading the dataset into RAM using standard computational tools –MATLAB, R, or Python. As already noted above, the next generation of neuroscience data must evolve beyond this paradigm.

---

[7] T.J. Sejnowski, P. S. Churchland, and J. A. Movshon, "Putting big data to good use in neuroscience," in *Nature Neuroscience*, vol. 17, no. 11, pp. 1440-1441, Nov. 2014.

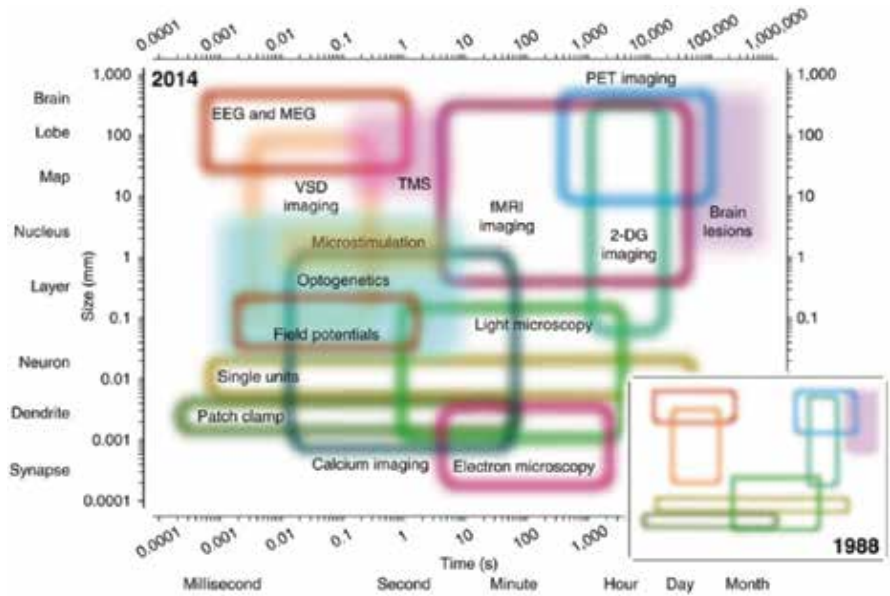

*Figure 2: Spatiotemporal scales of contemporary measurements of brain structure and function from Sejnowski et al., 2014.*

**Data Storage:** The data already being collected in modern neuroscience projects often exceeds the size of RAM, disk space on a local workstation, or even on a local data store. These data come from many different experimental modalities, spanning electron, visible light, and magnetic measurements. Today there are three key ways that neuroscientists collect data at such a scale. First, individual experiments can now lead to 100 TB of data, including serial electron microscopy and calcium physiology. Even MRI experiments can produce multiple TB of data products, including various intermediate stages of processing and formats. Second, the long tail of neuroscience includes many thousands of laboratories across the globe collecting data. Even if most of these experiments generate relatively small datasets, the amalgam of many of these datasets, collectively referred to as "mega-data", quickly reaches big data scales. Third, large-scale simulated data is not prominent yet in neurosciences. However, we are optimistic that soon the measured data from both large and small experiments will be sufficiently informative to justify detailed simulations which can result in huge amounts of data, just as the output of many physical simulations today are viewed as "big data."

These three different scenarios merit different solutions. To address large individual experiments we need scalable multidimensional spatial databases for neuroscience – for example the Open Connectome Project.[8] To amalgamate thousands of datasets from across the globe, we need neuroinformatics platforms to query and assemble such data. And for simulations we need to build detailed simulation models and possibly dedicated hardware.

---

[8] "Open Connectome Project." [Online] Available: http://www.openconnectomeproject.org/. [Accessed: April 21, 2015].

*Table I: List of useful brain sciences data resources*

| URL | Utility |
| --- | --- |
| http://openconnecto.me | Open science data & software |
| http://www.incf.org/resources/research-tools | Neuroinformatics tools |
| http://web.stanford.edu/group/brainsinsilicon/challenge.html | Dedicated neuromimetic hardware |
| https://wiki.humanconnectome.org/display/PublicData/Connecting+to+Connectome+Data+via+AWS | Commercial cloud storage solution for neuroscience |
| http://www.birncommunity.org/resources/data/ | Federal database of certain brain imaging data |
| http://crcns.org/data-sets | Home-grown brain data repository |
| http://www.loni.usc.edu/Software/Pipeline | Pipelining software for distributed computing |
| http://www.nature.com/sdata/ | Journal dedicated to publishing datasets |
| http://www.humanconnectome.org/data/ | Publicly shared dataset |
| http://fcon_1000.projects.nitrc.org/ | Consortia of publicly shared datasets |

**Data Access:** After data have been collected some means must be provided for others to access the data. There is already movement in this direction. Some groups are using commercial cloud solutions to make their data publicly available. Federal agencies have a strong history of archiving interesting public databases in the sciences, including in neurosciences. Other groups are building home grown databases with federal funding

All of these are promising approaches, but unification of access is essential. Just as cloud services are rapidly moving to a unified model, so also data access for science should be uniform, consistent, and easy to use. A key challenge is to manage the meta-data –provenance, content, experimental conditions, and so forth – that will be needed to correctly and accurately support data federation and analysis.

**Data Analysis:** Each area of science uses some methods of data analysis that are unique to that area, and others that are shared across the sciences more broadly. Some sciences, such as physics and climate research, rely mainly on highly specialized home-grown tools. This has also been the predominant model in neuroscience. However, while this approach was perhaps necessary in the past, it is not an effective mechanism for community-wide replicable science. It chains progress in the field to a patchwork of isolated tools that need constant maintenance and improvement. It also limits the speed with which new ideas, technologies, and tools are made available to the broader scientific community.

An important challenge and opportunity in neuroscience is to develop powerful, open-source tools that are broadly used and shared. The Galaxy Project[9] and the LONI Pipeline[10], for example, both provide promising examples of a new paradigm for replicable science based on shared data, common tools, and transparent analysis methodologies. Promoting open tools and methods for data analysis articulating the need and vision for these resources at a national level is a potential point of collaboration between computer science and brain sciences.

---

[9] "Galaxy Project." [Online] Available: http://galaxyproject.org/. [Accessed: April 20, 2015].
[10] "LONI Pipeline." [Online] Available: http://pipeline.bmap.ucla.edu/. [Accessed: April 20, 2015].

**Data Sharing:** The dogma of publishing a scientific paper that only reports highly distilled results is rapidly becoming outdated. The publication industry is combating this problem by creating scientific data journals, such as Nature Scientific Data and GigaScience. These journals, in turn, rely on the scientific data repositories. A particularly exciting possibility is co-locating or otherwise linking and archiving the data, data products and analysis tools for the data products. For example, the Human Connectome Project has already imaged over 500 subjects, and many labs from around the world have begun analyzing these data[11]. However, there is currently no place to publish tools and associated fine-grained results of these subsidiary analyses so that they can be shared and further refined. The Open Connectome Project is one example of an attempt to solve this issue by allowing analysts to post their results to annotation databases that are co-registered to the raw data, such that others can visualize, compare, and enhance the analyses of their peers.

Building on the evolving paradigms of sharing data, community sharing of tools and results in a format that promotes a new paradigm of rapid, well-structured, authenticated and archived data-derived publications will be essential for advancing the field.

## 5. Opportunities for Computing

The past half-century has seen momentous and accelerating progress in our understanding of the brain, while major research consortia are presently engaged in important efforts aimed at the complete mapping of brain structure and activity. And yet, despite all this excitement and progress, no overarching theory of the brain seems to be emerging. We believe that computational research can be productive in this connection. We thus close with a discussion of computing themes.

We suggest that the study of computing and the study of the brain interrelate in three ways, each suggesting a major research direction. First, as noted above, the experimental study of brain architecture and function is a massive-data problem. Making progress necessitates advances in computing and the realization of new computational tools. Second, the study of efficient algorithms and the design of intelligent autonomous systems should provide new ideas and inspiration concerning brain architecture and function. Finally, the remarkable efficiency (including energy efficiency) of the brain, once understood, may inspire radically new algorithmic or system organization approaches that could transform computing itself. Indeed, these are exactly the three bridges between brain science and computer science that we have included in Figure 1.

**A Computational Theory of the Brain:** The pioneers of both brain sciences and computer science – John von Neumann, Alan Turing, Walter Pitts, Warren McCulloch, David Marr, Herbert Simon – were very much aware of the relationship between the two areas, and they pursued it actively through the development of formal or philosophical computational theories. However, this important connection has been mostly stagnant over the past 30 years, that is, precisely at the time when the two fields exploded and therefore cross-fertilization could be most productive.[12] (One isolated exception was Leslie Valiant's work on "circuits of the mind", a rigorous computational approach to the cortex guided and informed by neuroscience findings.[13]) Is there a comprehensive computational theory that can inform our understanding of high-level brain function and the genesis of the mind? How would this theory be expressed and tested? What are the measurable "outputs" of the brain against which such a model could be validated?

**Machine Learning and the Brain:** Is there an ensemble of basic algorithmic ideas underlying the high-level function of the brain? We are nowhere near an answer to this question, but conjectures are starting to emerge. The success of machine learning, and of deep learning networks in particular, should be relevant to this quest. Yet the connections between the algorithms implemented in those systems and the operating principles of the brain remain unclear. Ongoing research seeks to rigorously explain the empirical success of machine learning. Such work often offers explanations based on the special structure and specialized distributions of the input.

---

[11] "Human Connectome Project." [Online] Available: http://www.humanconnectomeproject.org/. [Accessed: April 20, 2015].

[12] T. Poggio, "Marr's Approach to Vision." [online]. Available: ftp://publications.ai.mit.edu/ai-publications/pdf/AIM-645.pdf. [Accessed: April 20, 2015].

[13] L. G. Valiant, *Circuits of the mind*, Oxford University Press, 1994.

Connections between machine learning and the brain also go in the other direction: Current efforts for mapping the anatomy, structure, and function of the brain are hindered by conceptual and computational complexities, and a deluge of data. In comparison, mapping the human genome was trivial, because by the 1990s we knew a lot about the basic molecular mechanisms of life and we had a comprehensive, overarching theory of DNA function. What new advances in machine learning are needed to facilitate brain mapping research?

**Perception:** Our ability to act and react within the world is grounded by perception. Because we can control and observe sensory signals at the transducers, the processes of vision, touch, sound, taste and smell are some of the most accessible aspects of the brain. Furthermore, understanding how biological systems process sensory signals and optimally allocate resources for sensory processing may provide important clues for creating artificial systems that will operate in the natural world. What are the mathematical characteristics and latent structure of the distributions of the inputs the brain's sensors receive: that is, the environment within which the brain has evolved and developed?

**Language:** The path to the discovery of the fundamental algorithmic principles on which cortical computation is based may pass through language. Understanding the neural basis and cognitive structures of language, for example, provide a natural link between sensory behavior and cognitive function, or how cognitive process impact motor behavior through speech formation. If, for example, language emerged as a "last-minute adaptation," which arrived at a time when human cortex was essentially fully developed, language must have evolved in a way that takes full advantage of the brain's algorithmic principles. By studying what makes language so well adapted to our minds one could uncover important insights about the computational architecture of the brain.

**Lessons of the Brain:** Finally, the tremendous energy and computational efficiency of the brain, once understood, may inspire new ways and principles of organizing our computers and data centers.

# 6. The Future for Brain and Computing Research

Distilling the ideas above, we can see a number of challenges and opportunities for computer scientists working with brain scientists.

| Table 2: A summary of the Challenges and Opportunities | |
| --- | --- |
| **Challenge** | **Opportunity** |
| Extracting scientifically relevant information from the growing volume and variety of available images (from microscopy to MRI) remains a substantial challenge. In particular, reliable and scalable methods for neuroimage analysis and associated data analytics will be necessary to develop and test theories and simulations of the brain networks. | Developing new quantitative methods to create fine-grained models of neuroanatomy tied to computational functional simulation will provide new tools to study the effect of variation in brain structure to brain function, setting the stage for new insights on human development and disease processes. |
| Success in brain mapping, as well as in research on the brain-mind-body interface, requires the translation of a wide variety of raw signals into meaningful structures that can be shown to support causal models connecting structure to function to high-level behavior. | Creation of large databases of simultaneous recordings of behavioral and neuro-physiological data could allow machine learning based discovery of correlations. Such databases could also enable complementary clinical applications, e.g., predictions of upcoming health conditions. |
| Currently, it is impossible to connect brain models across scales (nanoscale to whole brain) and modalities (EM to MRI). Likewise, there are no methods to associate models across individuals or populations. | Create new modeling methods and scalable simulations to discover computational abstractions (motifs) at scale to accelerate progress in understanding the architecture of the brain. |
| New platforms supporting principled data federation, data analysis, and replicable science will radically alter the field of brain sciences. The platforms should be open-source and share not just data, but methods, results, and associated publications in order to accelerate the dissemination of both methods and results, and allow the broadest range of scientists access to the latest tools, insights, and results. | Conquering the greatest scientific problem of all time – understanding the brain – will absolutely require biologists, neuroscientists, psychologists, engineers, and computer scientists, working together. We need to educate undergraduate and graduate students and postdocs at the interface of neuroscience and computer science, including "immersion" experience and interaction with scientists from both disciplines. And we need to identify and promote career paths for researchers and educators at this interface. |
| Despite an explosion in brain-related data and new insights into small-scale structure and function and large-scale architecture of the brain, no overarching understanding of the brain's high-level function and the genesis of the mind appears to be emerging. In a somewhat related vein, despite impressive success over the past decade of brain-inspired machine learning algorithms, such as deep learning, no compelling connection has been made with the equally impressive success of the mammalian cortex. | The theoretical computer science research community has over the past decades developed productive and insightful models and incisive mathematical methodologies, which have been applied successfully to make progress in the sciences, including statistical and quantum physics, biology, and economics. Mobilizing this community around the exciting problems and opportunities in brain sciences – particularly the development of tools to model brain structure and activity at scale, will result in new ideas, insights, and progress in our understanding. |

**A Vision:** Where might this research take us? Consider the implications for neuroscience if we were to achieve innovations similar to what Google maps has done for cartography and navigation; what large scale finite element models and fine-scale sensor networks have done for weather forecasting; and how social network platforms running on the World-Wide Web have transformed social interaction.

*Cartography:* Compare the process of using a map 100 years ago with the process today. Last century, we carried paper maps with us. These often got torn, wet, or lost. They would be updated only when we purchased new ones, and even then only when the mapmakers published new maps. They did not tell us where we were, nor what direction we were going, nor anything about anybody else. Contrast that with navigating today using online mapping software. These maps tell us where we are, how to get where we want to go, how long it will take, and what we can expect along the way. They can tell us where our friends are, whether we've been there before, what other people thought of it. The maps are updated constantly, without our having to do anything about it. We can generate as many markers as we want, share them instantly and selectively with anybody in the world. And if we so desire the maps tell us where other people are as well, so we can avoid collisions. We can overlay topology, roads, buildings, and many other features. In short, current maps are dynamic, interactive, and multidimensional.

Neuroanatomy today is where global cartography was a century ago. We still publish books with hand drawn cartoons of neurological boundaries. Each book is dedicated to a particular species, with no obvious way of aligning the pages. What we need is a Neurocartography of the 21st century. This will include images of many different brains across spatiotemporal scales, spanning development and the evolutionary hierarchy. Each map will be linked to the others to enable overlaid views. Individuals will be able to annotate these maps with "reviews", including links to publications or direct links to analyses and results. These maps will be a reference point for every aspect of the neuroscience research enterprise. Before conducting a new experiment a neuroscientist will check the map. After conducting an experiment the neuroscientist will upload the result, creating either more map or more analyses of existing maps. We will be able to share links with our friends so they can follow us and always know where we are. Everything we do will be "neuro-tagged", just like everything we do today can be "geo-tagged".

*Simulation:* Geography is static, but weather is dynamic. Weather is shaped by geography, but the evolution of weather is a function of a complex web of factors driven by multiple sources of energy and the laws of physics. But we don't forecast weather by trying to model every molecule of air. Rather, weather is forecast by combining many sources of data within a complex mathematical model that approximates the physical dynamics of the atmosphere. As data becomes better, models become more precise computation more powerful, and predictions become better.

Imagine a future in which neurodynamics models can be formulated that are analogous to weather modeling and forecasting today. The starting point would be the unique properties of the neural system under study, informed by sensing and imaging modalities providing local temporal and/or spatial measurements of activity. A model analogous to a finite element model, but perhaps based on neural motifs or some abstraction of neural dynamics in a region of the brain, would be used to simulate neural or cognitive behavior into the future. At every time step the model could be compared to, or corrected by, updated measurements of brain activity. Inputs to the model would include stimuli – images, sounds, or other interactions – and outputs could be neural activity, motor activity, or even cognitive reporting of thought or sensation. As our models improve, our "forecasts" would get better, and as they get better, we would learn more about abstractions that describe the functioning of the brain. As this process scales, discrepancies in activity patterns would inform diagnosis of neural health, and modulation of activity due to therapy or drugs could be predicted. This in turn would lead to new and more precise ways to administer and manage interventions.

*Sharing Data:* Modern photo sharing sites allow one to see personal photos as well as photos taken by other people. Phototourism is a term that was coined to capture "visiting" physical places online using data shared by others. In the future this will happen in the brain sciences. As noted above, neurocartography provides a frame of reference for thinking about "places I want to visit" within the brain. But, unlike a physical monument, every brain is different. Thus, to really understand some aspect of the brain, we will not only need to see every "picture" (image, time-series recording, etc.), but we will also need to understand the surrounding context – age, gender, health condition, relevant stimuli, etc. We will need "viewers" that are really sophisticated analytical engines that let one "compute forecasts," ask questions and derive answers without building analytics from scratch. Finally, we will need a way to publish that links back to these data archives and allows others to replicate the results independently, and refine them as new models and theories arise.

These analogies are, by their nature, a coarse effort to evoke ideas for a possible future. However, it is almost certain similar ideas will emerge and become essential elements for brain research. Implicit in all are ideas that span the four major thrusts of this workshop: brain mapping, connecting mind, brain and body, computation, and data.

## 7. Computing and Brain Sciences: Re-establishing a Joint Destiny

The 80 participants of the workshop, computer scientists and brain researchers with strong research interests in both fields, spent two days debating the state of the art in brain sciences, and admiring the essential and diverse ways in which it relates to research in computer science. We left the workshop convinced that these two key disciplines are destined to work hand-in-hand in the coming decades to address the grand challenges in the research at their interface, and to create a common culture shared by the researchers working on both disciplines.

# Participant List

| First Name | Last Name | Affiliation |
|---|---|---|
| Charles | Anderson | Colorado State |
| Sanjeev | Arora | Princeton |
| Satinger | Baveja Singh | Michigan |
| Krastan | Blagoev | National Science Foundation |
| Matt | Botvinick | Princeton |
| Randal | Burns | JHU |
| Jose | Carmena | Berkeley |
| Miyoung | Chun | Kavli |
| Sandra | Corbett | CRA |
| Christos | Davatzikos | UPenn |
| Susan | Davidson | UPENN |
| Heather | Dean | National Science Foundation |
| Jim | Deshler | National Science Foundation |
| Ann | Drobnis | CCC |
| Jim | Duncan | Yale |
| Tatiana | Emmanouil | CUNY |
| Greg | Farber | OTDC/NIMH/NIH |
| Naomi | Feldman | Maryland |
| Vitaly | Feldman | IBM |
| Charless | Fowlkes | UC Irvine |
| Jeremy | Freeman | HHMI |
| Jack | Gallant | Berkeley |
| Shafi | Goldwasser | MIT |
| Polina | Golland | MIT |
| Greg | Hager | JHU |
| Dan | Hammerstrom | DARPA / MTO |
| Jim | Haxby | Dartmouth |
| Sean | Hill | Human Brain Project |
| Vasant | Honavar | Penn State |
| Bill | Howe | U. Washington |
| Sabrina | Jacob | CRA |
| Bala | Kalyanasundaram | National Science Foundation |
| Nancy | Kanwisher | MIT |
| Konrad | Koerding | Northwestern |
| S. Rao | Kosaraju | NSF |
| Yann | Lecun | NYU / Facebook |
| Richard | Lewis | UMich |

| First Name | Last Name | Affiliation |
|---|---|---|
| Jeff | Lichtman | Harvard |
| Deborah | Lockhart | National Science Foundation |
| Chris | Martin | The Kavli Foundation |
| Brandon | Minnery | IARPA |
| Sandro | Mussa-Ivaldi | Northwestern |
| Shelia | Nirenberg | Cornell |
| James | Olds | National Science Foundation |
| Aude | Oliva | MIT |
| Bruno | Olshausen | UC Berkeley |
| Christos | Papadimitriou | Simons |
| Pietro | Perona | CalTech |
| Hanspeter | Pfister | Harvard |
| Tal | Rabin | IBM |
| Rajesh | Rao | U. Washington |
| Giulio | Sandini | Italian Institute of Technology |
| Stefan | Schaal | USC |
| Nicolas | Schweighofer | USC |
| Terry | Sejnowski | Salk Institute |
| Fei | Sha | University of Southern Califonria |
| Maryam | Shanechi | University of Southern California |
| Hava | Siegelman | UMass-Amherst |
| Ambuj | Singh | UC Santa Barbara |
| James | Smith | Wisconsin |
| Sara | Solla | Northwestern University |
| Sharif | Taha | Kavli Foundation |
| Josh | Tenenbaum | MIT |
| Bertrand | Thirion | Neurospin |
| Paul | Thompson | University of Southern California |
| Francisco | Valero-Cuevas | University of Southern California |
| Les | Valiant | Harvard |
| Helen | Vasaly | CCC |
| Santosh | Vempala | Georgia Tech |
| Ragini | Verma | UPenn |
| Joshua | Vogelstein | Johns Hopkins University |
| Howard | Wactlar | Carnegie Mellon University |
| Kenneth | Whang | National Science Foundation |
| Ross | Whitaker | Utah |
| Martin | Wiener | National Science Foundation |
| Rebecca | Willett | Wisconsin |

**Notes:**

_______________________________________________________________________

_______________________________________________________________________

_______________________________________________________________________

_______________________________________________________________________

_______________________________________________________________________

_______________________________________________________________________

_______________________________________________________________________

_______________________________________________________________________

_______________________________________________________________________

_______________________________________________________________________

_______________________________________________________________________

_______________________________________________________________________

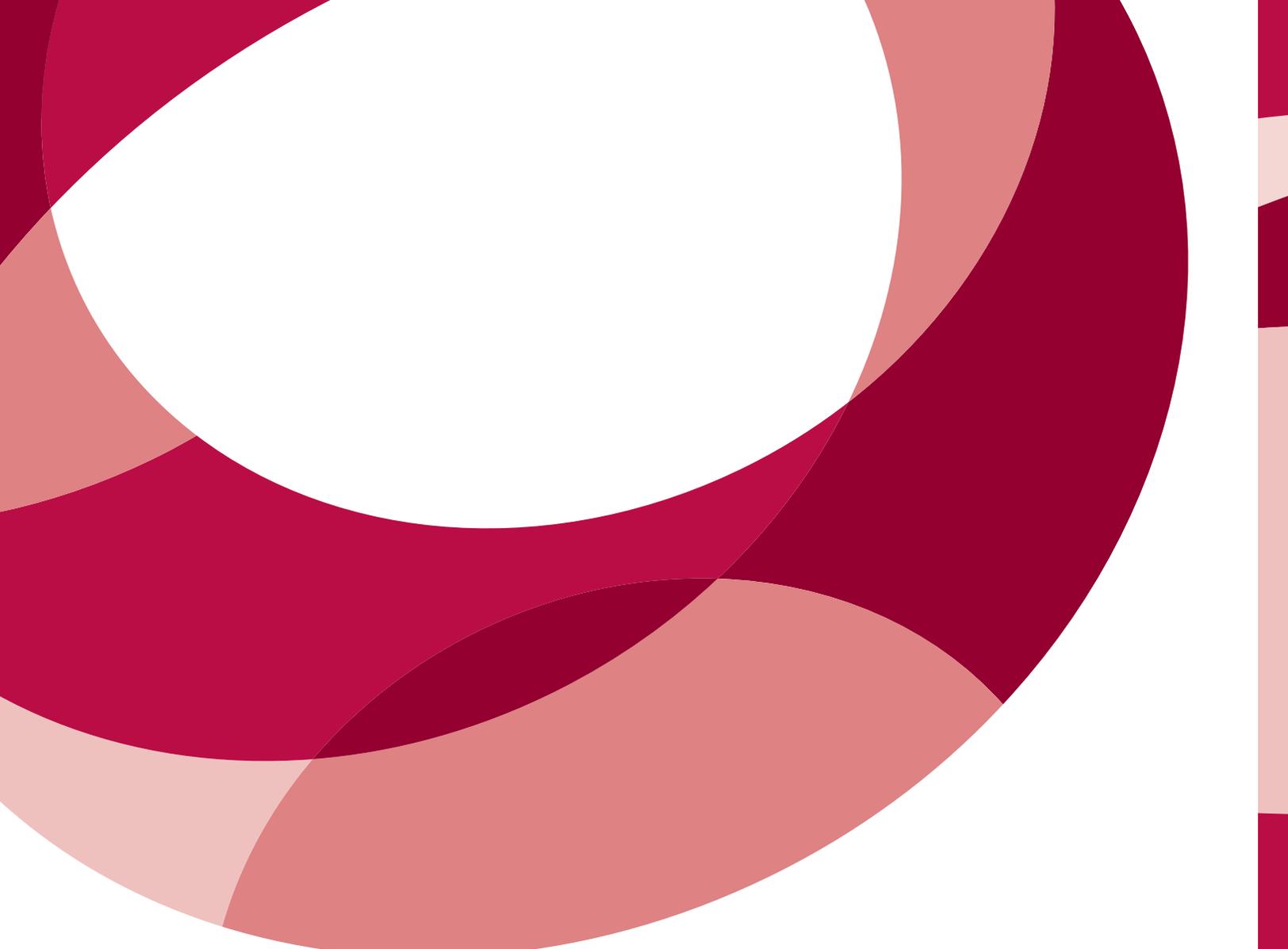

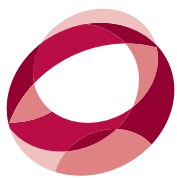

**CCC**
Computing Community Consortium
Catalyst

1828 L Street, NW, Suite 800
Washington, DC 20036
P: 202 234 2111 F: 202 667 1066
www.cra.org cccinfo@cra.org